\documentclass[modern]{aastex63}

\usepackage{graphicx}
\usepackage{epstopdf}
\usepackage{ulem} 
\usepackage{color} 
\usepackage{lineno}
\usepackage{amsmath}
\usepackage{lipsum} 

\def\lsim{\;\raise0.3ex\hbox{$<$\kern-0.75em\raise-1.1ex\hbox{$\sim$}}\;}
\def\gsim{\;\raise0.3ex\hbox{$>$\kern-0.75em\raise-1.1ex\hbox{$\sim$}}\;}

\definecolor{purple}{RGB}{200,100,255} 

\newcommand{\dNdp}{\mathrm{d} N/\mathrm{d} p}

\newcommand{\delthtmax}{\delta\theta_\mathrm{max}}

\newcommand{\GamZ}{\Gamma_0}

\newcommand{\betaZ}{\beta_0}

\newcommand{\xx}[1]{\!\times\!10^{#1}}

\newcommand{\epsB}{\epsilon_{B}}

\newcommand{\lambdamfp}{\lambda_\mathrm{mfp}}
\newcommand{\etamfp}{\eta_\mathrm{mfp}}

\newcommand{\pmax}{p_\mathrm{max}}

\newcommand{\gb}{\Gamma\beta}
\newcommand{\pipl}{\vec{\pi}_\mathrm{pl}}
\newcommand{\phiMC}{\vec{\phi}_\mathrm{MC}}
\newcommand{\pemp}{\vec{p}_\mathrm{emp}}

\newcount\listnorom
\newcommand\listromanDE{\global\advance \listnorom by 1
{\lowercase\expandafter{(\romannumeral\listnorom)}\ }}

\newcount\listnumber
\newcommand\listDE{\global\advance \listnumber by 1
{\lowercase\expandafter{(\number\listnumber)}\ }}

\newcount\Inum
\newcount\IInum
\newcount\IIInum
\newcount\IVnum
\def\I{\global\multiply\IInum by 0 \global\multiply\IIInum by 0
            \global\multiply\IVnum by 0 \global\advance \Inum by 1
            {\the\Inum. }}
\def\II{\global\multiply\IIInum by 0\global\multiply\IVnum by 0
       \global\advance \IInum by 1 {\the\Inum.\the\IInum. }}
\def\III{\global\multiply\IVnum by 0\global\advance \IIInum by 1
            {\the\Inum.\the\IInum.\the\IIInum. }}
\def\IV{\global\advance \IVnum by 1
            {\the\IVnum. }}
\shorttitle{Max electron energy in DSA}
\shortauthors{Warren et al.}

\begin{document}

\title{The maximum energy of shock-accelerated electrons in a microturbulent magnetic field} 

\vskip24pt

\newcommand{\iTHEMS}{RIKEN Interdisciplinary Theoretical and Mathematical Sciences Program (iTHEMS), Wak\={o}, Saitama, 351-0198 Japan}
\newcommand{\ABBL}{Astrophysical Big Bang Laboratory (ABBL), RIKEN Cluster for Pioneering Research, Wak\={o}, Saitama, 351-0198 Japan}
\newcommand{\ryerson}{Department of Physics, Ryerson University, Toronto, Canada, M5B 2K3}
\newcommand{\purdue}{Department of Physics and Astronomy, Purdue University, West Lafayette, IN 47907-2036, USA}
\newcommand{\IARAS}{Institute of Astronomy, Russian Academy of Sciences, Moscow, 119017 Russia}

\author[0000-0002-3222-9059]{Donald~C. Warren}%
\affiliation{\iTHEMS}
\author[0000-0003-0599-0069]{Catherine~A.~A. Beauchemin}%
\affiliation{\ryerson}
\affiliation{\iTHEMS}
\author[0000-0002-0960-5407]{Maxim~V. Barkov}%
\affiliation{\purdue}
\affiliation{\ABBL}
\affiliation{\IARAS}
\author[0000-0002-7025-284X]{Shigehiro Nagataki}%
\affiliation{\ABBL}
\affiliation{\iTHEMS}

\begin{abstract}

Relativistic shocks propagating into a medium with low magnetization are generated and sustained by small-scale but very strong magnetic field turbulence.  This so-called ``microturbulence'' modifies the typical shock acceleration process, and in particular that of electrons.  In this work we perform Monte Carlo (MC) simulations of electrons encountering shocks with microturbulent fields.  The simulations cover a three-dimensional parameter space in shock speed, acceleration efficiency, and peak magnetic field strength.  From these, a Markov Chain Monte Carlo (MCMC) method was employed to estimate the maximum electron momentum from the MC-simulated electron spectra.  Having estimated this quantity at many points well-distributed over an astrophysically relevant parameter space, an MCMC method was again used to estimate the parameters of an empirical formula that computes the maximum momentum of a Fermi-accelerated electron population anywhere in this parameter space.  The maximum energy is well-approximated as a broken power-law in shock speed, with the break occurring when the shock decelerates to the point where electrons can begin to escape upstream from the shock.

\end{abstract}

\section{Introduction}
\label{sec:intro}

When particles are accelerated by the Fermi process at a collisionless shock, the distribution takes the form of a power law in momentum, with an exponential cutoff. While energy gains from acceleration exceed losses, the distribution is that of a power law; above some momentum, however, energy losses dominate and the number of particles drops off exponentially. The resulting function may be written
\begin{equation}
  \frac{\mathrm{d} N}{\mathrm{d} p} = K p^{-\sigma} \mathrm{e}^{-(p/\pmax)^{\chi}} .
  \label{eq:pl_def}
\end{equation}
In the above equation, $K$ is a normalization constant, $\sigma$ is the spectral index of the power law, $p_\mathrm{max}$ is the location of the exponential cutoff, and $\chi$ controls the curvature of the turnover.  In the case of Bohm diffusion limited by a free escape boundary, $\chi = 1$ \citep{CBA2009}.  If particle energies are instead limited by radiative losses such that $\dot{p} \propto p^{2}$, then $\chi = 2$ \citep{ZirakashviliAharonian2007,LKA2012}.  In the extreme limit $\chi \gg 1$, the exponential turnover behaves like a step function.  An additional component is expected in physical distributions, that of the particles that crossed the shock but were not further accelerated.  This ``thermal'' population is discussed elsewhere in the literature \citep{GianniosSpitkovsky2009,WEBL2015,ResslerLaskar2017}, but since it occurs at the low-energy end of the particle distribution we ignore it henceforth.  For this work we are not interested in how particles are injected into the acceleration process; we focus solely on what happens to the particles once they are being Fermi-accelerated.

Both $\sigma$ and $\pmax$ can be predicted analytically under certain simplifying conditions.  If particle orientations diffuse isotropically, and if the shock is assumed to be discontinuous, then $\sigma$ depends only on the shock Lorentz factor $\GamZ$; one can then recover $\sigma = 2$ in the non-relativistic limit $\GamZ \rightarrow 1$ and $\sigma \approx 2.2$ in the ultra-relativistic limit $\GamZ \rightarrow \infty$ \citep{KeshetWaxman2005}.  This result has been confirmed by many numerical studies both pre- and post-dating the analytical work \citep{Ostrowski1988, ERJ1990, EllisonDouble2002, Baring2004, AMN2008, EWB2013}.

To determine $\pmax$ one must assume an energy loss mechanism.  If radiative losses such as synchrotron (or synchrotron self-Compton) are responsible, then one may set the acceleration time equal to the loss time \citep{GallantAchterberg1999, Achterberg_etal_2001, Aharonian_etal_2002, LemoinePelletier2003, PiranNakar2010, Bykov_etal_2012SSRv, Lemoine2013}.  The resultant formula for $\pmax$ has many more dependencies than that for $\sigma$; as shown in the cited works $\pmax$ potentially depends on large-scale hydrodynamical parameters (e.g. shock speed) and microphysical parameters (e.g. magnetic field strength, turbulence characteristics, or scattering model).

Particle-in-cell (PIC) simulations \citep{Nishikawa_etal_2007, Spitkovsky2008, SSA2013, Ardaneh_etal_2015, ACN2016} offer a first-principles method for determining the distributions particles take as they interact with a collisionless shock.  By self-consistently treating both how particles self-generate magnetic field turbulence and how they scatter in it, PIC simulations act as the ``gold standard'' for numerical studies.  However, they are extremely computationally intensive, and this severely limits their ability to simulate the times and volumes needed to discuss the value of $\pmax$ in realistic settings. \footnote{The simulations presented in \citet{Spitkovsky2008} and \citet{SSA2013} do show a power law of electrons forming as a result of shock acceleration, but the power law spans less than an order of magnitude in energy.  An electron distribution accelerated at a collisionless shock may cover $5-6$ orders of magnitude; a back-of-the-envelope estimate suggests that PIC simulations like those cited would need to run for $10^{4}$, or more, times longer---with a correspondingly larger simulated volume---to approach a power law of such breadth.}

While PIC simulations can not yet address the value of $\pmax$, they are in general agreement about the structure of the magnetic field around the relativistic shocks they simulate.  Such shocks are susceptible to a wide variety of plasma instabilities \citep{MedvedevLoeb1999, BGD2010, LemoinePelletier2011}, and these instabilities drive small-scale ``microturbulence'' in the magnetic field that can exceed 10\% of the local energy density \citep[e.g.][]{SSA2013}.  This microturbulence decays rapidly behind (and also in front of) the shock, meaning particles probe magnetic fields of different strengths and turbulence spectra.  The maximum momentum attainable by particles can still be determined by equating acceleration time and loss time, but both of those quantities are more complicated than they are in a zeroth-order model \citep{Lemoine2013}.

Monte Carlo simulations \citep{Ostrowski1988, ERJ1990, EllisonDouble2002, LemoinePelletier2003, Baring2004, NiemiecOstrowski2006, EWB2013, WEBL2015} trade additional assumptions about the shock microphysics for substantially lower computational costs. Instead of handling magnetic field turbulence and particle scattering self-consistently, Monte Carlo schemes typically fix one or both.  The computational cost of Monte Carlo codes is sufficiently low that they can be used to probe $\pmax$, albeit sparsely, over a meaningful range of shock parameters and magnetic field configurations.

In this work, we apply an established Monte Carlo code \citep{Ellison1985, ERJ1990, EllisonDouble2002, VBE2008, EWB2013, WEBL2015, WEBN2017} to the problem of particle acceleration in the microturbulent field.  We simulate electron acceleration, and radiative losses, over a multidimensional physical parameter space.  From the accelerated particle spectra thus obtained, we identify an empirical formula that quickly computes $\pmax$ over a wide parameter space of astrophysical relevance.

\section{Numerical methods}
\label{sec:code}

The Monte Carlo (hereafter MC) model used here assumes a steady-state shock and explicitly defines how particles scatter in the magnetic turbulence, in effect solving the Boltzmann equation with collective scattering \citep{EllisonEichler1984, EWB2013}.  This steady-state approximation is valid as long as the particle acceleration time is shorter than the dynamical time of the system.

Particle motions, energy losses, and scatterings are treated in a series of discrete time steps $\delta t$.  The MC model assumes that particles scatter with a mean free path proportional to their gyroradius $r_{g} = \gamma m v c / (qB)$ in the local magnetic field $B$,
\begin{equation}
  \lambdamfp = r_{g} \etamfp
  \label{eq:mfp}
\end{equation}
where $\etamfp$ characterizes the strength of the scattering; the case of $\etamfp = 1$ is called Bohm diffusion.  Bohm diffusion is commonly assumed because of its simplicity, but there is no conclusive evidence that $\etamfp$ has any particular value, or indeed a single value at all.  Observations of rapid variability in bright knots of the supernova remnant RX J1713.7-3946 suggest that particle acceleration takes place with $\etamfp \sim 1$ \citep{Uchiyama_etal_2007}.  In the ultrarelativistic limit, interpreting GeV emission in the afterglows of gamma-ray bursts (GRBs) as due to synchrotron emission places strict limits of $\etamfp \lesssim 1$ on the acceleration process \citep{SagiNakar2012}.  Simulations of particle scattering in the relativistic regime variously suggest that $\etamfp \gsim 10$ \citep{LemoineRevenu2006} or $\etamfp \sim 1$ \citep{RiordanPeer2019}.  Indeed, the discussion surrounding $\etamfp$ is far from resolved (and the dependence of $\lambdamfp$ on $r_{g}$ may not be linear in some regimes \citep{PPL2011}), and for simplicity Equation~\ref{eq:mfp} is assumed to hold everywhere in the shock structures simulated.

The MC model divides each gyroperiod into $N_{g}$ smaller propagation and scattering steps which take place over a time $\delta t$.  That is, $\delta t \propto r_{g}/N_{g}$.  The amount by which particle orientations are allowed to change at each scattering event depends on both $\etamfp$ and $N_{g}$ \citep{EWB2013}: $\delthtmax \propto (\etamfp N_{g})^{-1/2}$.  In the continuous limit of $\delta t \rightarrow 0$ (equivalent to $N_{g} \rightarrow \infty$), we approach pitch-angle diffusion as used in \citep{KeshetWaxman2005}.  The value of $N_{g}$ used in the simulations was tested to confirm that larger values did not affect the results.

The only energy gain process used in the MC model is first-order Fermi acceleration, i.e. crossing the shock and scattering off of turbulence in the new rest frame.  There are numerous other mechanisms by which particles in the vicinity of a shock could gain energy, including wake field acceleration \citep{Tajima_Dawson_1979}, second-order Fermi acceleration \citep[][and references therein]{Virtanen_Vainio_2005}, and cross-shock potentials \citep[e.g.,][]{Tran_Sironi_2020}. We assume that once injection into the first-order Fermi process occurs, those energy gains dominate those from other sources.

The code can treat the acceleration process of particles with any charge to mass ratio, and has in the past been applied to electrons, protons, and heavier ions.  For this work we restrict ourselves to considering only electrons.  Electrons carry significantly less momentum and energy flux than do ions, and their ability to influence the shock structure is similarly limited.  Even in relativistic shocks with $\GamZ \sim 30$, nonlinear interactions between the shock and the accelerated particles can change the velocity profile of the shock \citep{EWB2013}.  At higher particle energies, especially in the limit of relativistic shocks, the induced precursor is of sufficiently small extent that it is ignorable.  Since the primary objective of this work is to estimate the value of $\pmax$ for the electron spectrum given in Equation~\ref{eq:pl_def}, nonlinear effects can be reasonably neglected.

Another break from previous work with this code is the magnetic field profile used.  Non-relativistic versions of the code self-consistently handle generation and decay of magnetic field turbulence \citep{VBE2008, VBE2009, Bykov_etal_2014}, but the relativistic version assumed that the mean field was parallel to the plane of the shock and of uniform strength throughout the shock structure (both up- and downstream from the shock).  In the work presented here we assume that the magnetic field is dominated by isotropic turbulence, consistent with the downstream region in PIC simulations. The magnetic field upstream is more ordered due to the filamentation instability, but in relativistic shocks electrons spend the bulk of their time scattering in the downstream medium.  The strength of the magnetic field is related to the local energy density by the parameter $\epsB$.  The peak value of both magnetic field and $\epsB$ occurs at the shock, and may be expressed in terms of bulk hydrodynamical parameters,
\begin{equation}
  \epsilon_\mathrm{B,pk} = \frac{ B^{2} }{ 16 \pi \GamZ^{2} n_{0} m_{p} c^{2} } ,
  \label{eq:epsB_pk}
\end{equation}
where $B$ is the mean local magnetic field, $\GamZ$ is the Lorentz factor of the shock, and $n_{0}$ is the rest-frame ambient density the shock is presently encountering \citep[cf. the definition of proper energy density in][]{BlandfordMcKee1976, GranotSari2002}.

Upstream and downstream from the shock, the magnetic field used is consistent with PIC simulations \citep{Keshet_etal_2009, SSA2013} and previous analytical work \citep{Lemoine2013}:
\begin{equation}
  \epsB(x) = \epsilon_{B,\mathrm{pk}}
  \begin{cases}
    10.4 \left|x\right|^{-0.6}  &  x < -50  \\
    1  &  -50 \leq x \leq 50  \\
    50 x^{-1}  &  50 < x
  \end{cases}
  \label{eq:epsB_form}
\end{equation}
in which the distance $x$ from the shock is measured in ion skin depths, $\lambda_\mathrm{sd} = (\GamZ m_{p} c^{2} / 4 \pi e^{2} n_{0})^{1/2}$.  This functional form has a plateau within 50 plasma skin depths of the shock, then decays upstream (negative positions) and downstream (positive positions).  The plateau has been used in previous analytical treatments of microturbulence in GRB afterglows \citep{Lemoine2013}, but is a simplification of the sharper, exponential-like peak in $\epsB$ found in PIC simulations.  The downstream decay rate of $x^{-1}$ is expected from PIC simulations \citep{CSA2008} and interpretations of GRB afterglows with microturbulence \citep{LLW2013}. The upstream decay rate matches the precursors presented in \citet[][Fig.~7]{SSA2013}; we have not seen an extended discussion of this decay rate in the literature, so we do not make any claims of universality here.

The oblique instability responsible for the microturbulence is expected to disappear when the shock speed drops below a certain critical value, $\GamZ \sim 10$ for reasonable choices of physical parameters \citep{LemoinePelletier2011, Pelletier_etal_2017}. The calculation assumes a relativistic shock, however, and it is less clear what instabilities are present in the mildly relativistic regime.  At low Lorentz factors both analytical and numerical approaches encounter difficulty, and there are comparatively fewer studies in the literature.  Numerical work on mildly relativistic shocks suggests that short-wavelength turbulence exists in the shock precursor at Lorentz factors as low as $\GamZ = 1.5$ \citep{RBO2018}.  We therefore assume here that amplified microturbulence is present at all Lorentz factors considered, and remain agnostic as to the precise source of the turbulence.

As particles scatter in the above-mentioned magnetic field turbulence, they experience radiative losses.  We consider only synchrotron losses in this work.  Electrons should additionally radiate away energy via the inverse Compton process, and in the strong magnetic fields present near the shock synchrotron self-Compton (SSC) can be significant \citep{SariEsin2001}.  The rate of SSC losses depends on the synchrotron spectrum at an electron's location within the shock structure, which in turn depends on the local electron distribution.  The full calculation of synchrotron-SSC cooling is thus highly nonlinear. \citet{SBM2010} discuss the problem of accelerating electrons that are simultaneously suffering synchrotron and SSC losses, but their work covers only interactions in the Thomson regime.  As magnetic field strengths and electron energies increase, the Klein-Nishina cutoff on SSC cross-section becomes relevant and serves to suppress SSC losses.  SSC in the Klein-Nishina regime has been considered by many authors, but their work assumed an injected electron distribution rather than calculating it self-consistently \citep{NAS2009, Wang_etal_2010, BDBK2012}.  We are unaware of an attempt to simultaneously treat electron acceleration and SSC cooling in the Klein-Nishina regime, and so we also ignore these complications to focus instead on the much simpler synchrotron loss process.

We include another means of limiting particle energy in the code: a free escape boundary (FEB) upstream of the shock.  Any particles reaching the boundary are considered to decouple from the shock, and are removed from the acceleration process.\footnote{As pointed out by \citet{Drury2011}, free escape boundaries are in tension with the assumptions of a planar, stationary shock used here.  However, the focus here is purely on electrons, which remain coupled to the shock for a shorter period of time than do ions, and so the shock does not evolve much while the electrons are being accelerated.  Further, the separation of the FEB and the shock is much smaller than the shock radius for the parameter space used here; particles confined by these FEBs would see the shock as planar.}  Particle escape from a shock can only be identified after the fact; particles scattering ahead of a shock have a non-zero probability of returning back to the shock regardless of how far from the shock they get, and this probability is increased with an expanding shock \citep{Drury2011}.  An additional complication arises since particle diffusion lengths increase in the weaker magnetic fields far upstream from the shock.  Rather than track all particles indefinitely, or introduce a random chance of escape between each scattering event, we approximate particle escape with the use of a sharp boundary.  The FEB acts as an additional limitation on particle energy since diffusion lengths are proportional to energy (Equation~\ref{eq:mfp}).  The location of the FEB for all MC simulations is
\begin{equation}
  x_\mathrm{FEB} = -10^{3} \frac{\GamZ \, \betaZ \, m_{p} c^{2} }{ e \, B_{0} }
  \label{eq:FEB}
\end{equation}
where the negative sign denotes a position upstream from the shock, and $\betaZ = u_{0}/c = \sqrt{1 - \GamZ^{-2}}$. That is, the physical location of the FEB depends on both the shock speed and the ambient magnetic field $B_{0}$, which taken here to be $10^{-5}$~G.  Beyond setting the physical scale of the FEB, the ambient magnetic field is largely irrelevant because it is dominated everywhere in the MC simulations by the amplified turbulence.

It is easier for particles to scatter out to the FEB in mildly relativistic shocks for several reasons.  First, Equation~\ref{eq:FEB} shows that the physical location is closer to the shock when $\GamZ$ is lower.  Second, according to Equation~\ref{eq:epsB_pk} the magnetic field is weaker around slower shocks.  This means longer mean free paths per Equation~\ref{eq:mfp}, so the same physical distance represents fewer diffusion lengths when the magnetic field is reduced.  The weaker field also means that electrons experience smaller radiative losses; they therefore keep more of their energy as they scatter towards the FEB and can make the trip in fewer scattering steps.  Taken as a whole, if the FEB acts to limit the maximum electron energy we expect that maximum to be an increasing function of $\GamZ \betaZ$; once electrons can no longer reach the FEB, the relationship between shock speed and maximum energy will change.

All shocks simulated in this work assume that acceleration occurs in the test-particle limit; that is, the presence of accelerated particles upstream from the shock does not modify the velocity profile.  In actuality, pressure from these accelerated particles acts to slow down inflowing plasma, generating a shock precursor via a nonlinear feedback loop.  This nonlinear shock modification occurs at all shock speeds, from nonrelativistic to ultrarelativistic \citep{EllisonEichler1984,RKD2009,EWB2013}.  However, the precursor in a relativistic shock is weak compared to those of non-relativistic shocks \citep{EWB2013}; so while nonlinear effects may be present, their treatment is deferred to future work.

\section{Estimating $\pmax$ from the Monte Carlo results}
\label{sec:results}

The MC code described above was used to simulate particle acceleration at numerous points in a three-dimensional parameter space.  The proper speed of the shocks, $\gb$, was allowed to vary between 0.83 (corresponding to $\GamZ = 1.3$) to 340.  The plateau value of $\epsB$ in the vicinity of the shock ranged between $10^{-4}$ and $10^{-1}$.  The acceleration efficiency $\etamfp$ varied between 1 and 10.  The strongest magnetic fields in the simulations occur at the shock.  Their strength is, by the definition of $\epsB$, a product of that parameter and the local energy density:
\begin{equation}
  B = \left( 16 \pi n_{0} \GamZ^{2} m_{p} c^{2} \epsB \right)^{1/2}
  \label{eq:B_pk_strength}
\end{equation}
This strength depends on both $\gb$, by way of $\GamZ^{2}$, and on the product $\epsB n_{0}$ (where $n_{0}$ is the ambient density of the material being swept up by the shock). Each individual MC run can thus be characterized by a parameter triplet $\phiMC = (\gb, \epsB n_{0}, \etamfp)$.

Over this wide, physically-relevant range of $\phiMC$ parameters, the MC code was used to simulate 20 iterations (with a different initial random number seed) for each parameter triplet.  From the resulting 20 electron spectra, we estimated a posterior likelihood distribution (PLD) of the parameters $\pipl = (\log_{10} \pmax, \sigma, \chi, \log_{10} K)$ via Markov chain Monte Carlo (MCMC), as detailed in Appendix~\ref{sec:MCMC_details}.

\begin{figure}
  \includegraphics[width=0.5\columnwidth]{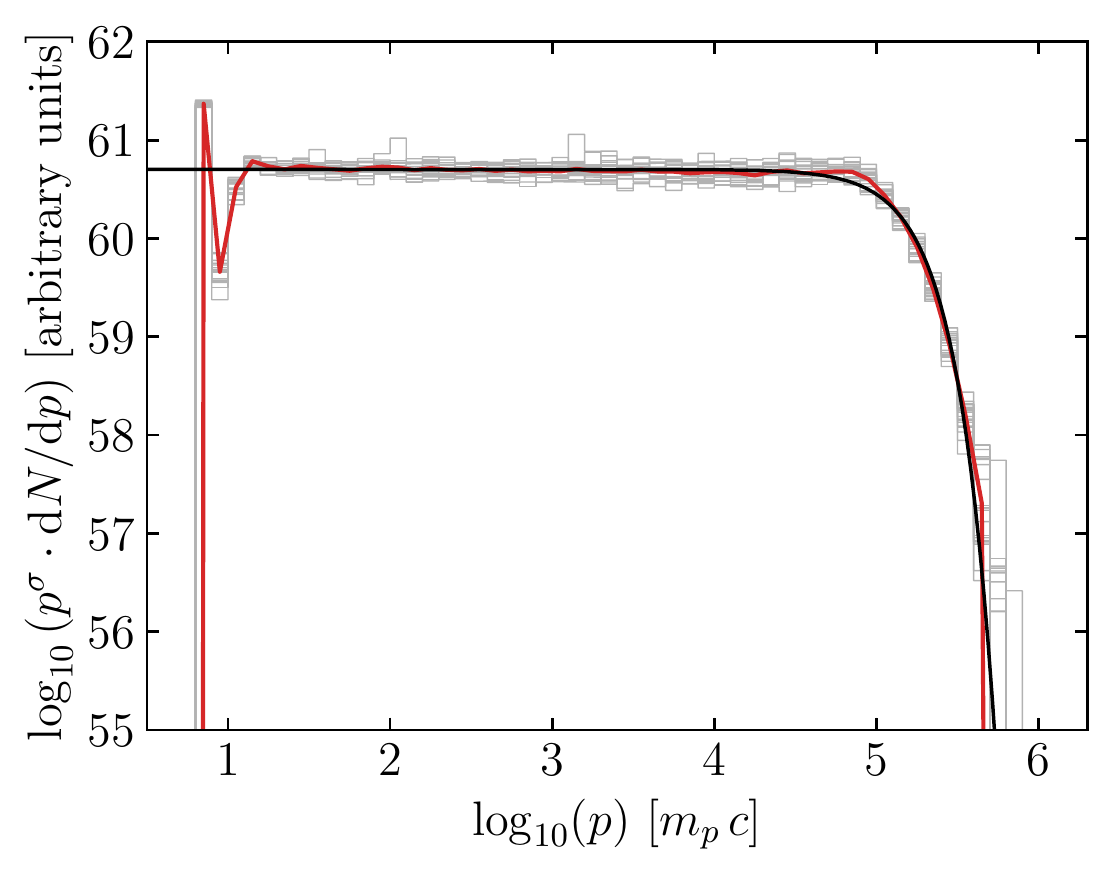}
  \caption{Comparison between the electron spectra obtained from the MC code, and the spectrum described by Equation~\ref{eq:pl_def} for the MCMC-identified highest-likelihood $\pipl$ (see Section~\ref{sec:fit}).  Histograms of electron spectra from 20 different MC iterations (using the same $\pipl$ but different random number seeds) are shown with light grey lines.  Their geometric average is plotted in red.  The black curve is Equation~\ref{eq:pl_def} using the highest-likelihood parameters identified via MCMC.  The input parameters for the MC simulation were $\phiMC = (\gb = 100, \etamfp = 1, \epsB n_{0} = 10^{-4})$. The corresponding highest-likelihood parameters estimated by MCMC were $\pipl = (\log_{10} \pmax = 5.13, \sigma = 2.18, \chi = 1.86, \log_{10} K = 60.7)$.}
  \label{fig:mc_results}
\end{figure}
An example of this process is given in Figure~\ref{fig:mc_results}, showing both the MC realizations and the MCMC-estimated spectrum.  The MC electron distribution has two components: a ``thermal peak'' (composed of electrons that crossed the shock but did not enter the acceleration process) and the non-thermal tail (made up of electrons that did shock-accelerate).  The sharpness of the thermal peak is an artifact of how the MC code handles energy gain upon shock crossing\footnote{Specifically, particles gain energy based on the velocity difference between their initial and final positions between scatterings.  For cold particles encountering a relativistic shock, the energy gain is much greater than the initial thermal energy the particles possessed.  The MC code does not try to artificially redistribute energy to recover a thermal distribution, but PIC simulations do show that such a distribution exists in the downstream region of the shock \citep[e.g.,][]{SSA2013}.}; it does not impact the acceleration process that produces the non-thermal tail.  The non-thermal distribution is well-captured by Equation~\ref{eq:pl_def}, including the extended power-law tail and the exponential turnover at $\pmax$.

The momentum value of the thermal peak in Figure~\ref{fig:mc_results} is a free parameter in this work.  Its position can be fixed using either a PIC simulation or by assuming a specific fraction of energy placed in the electron distribution ($\epsilon_{e}$ in the literature), but we are not interested here in the low-energy portion of the spectrum.  We instead place the thermal peak at a sufficiently small momentum that the power law generated by shock acceleration can cover several decades, allowing us to adequately estimate the $\sigma$ parameter in Equation~\ref{eq:pl_def}.

\begin{figure*}
  \includegraphics[width=\textwidth]{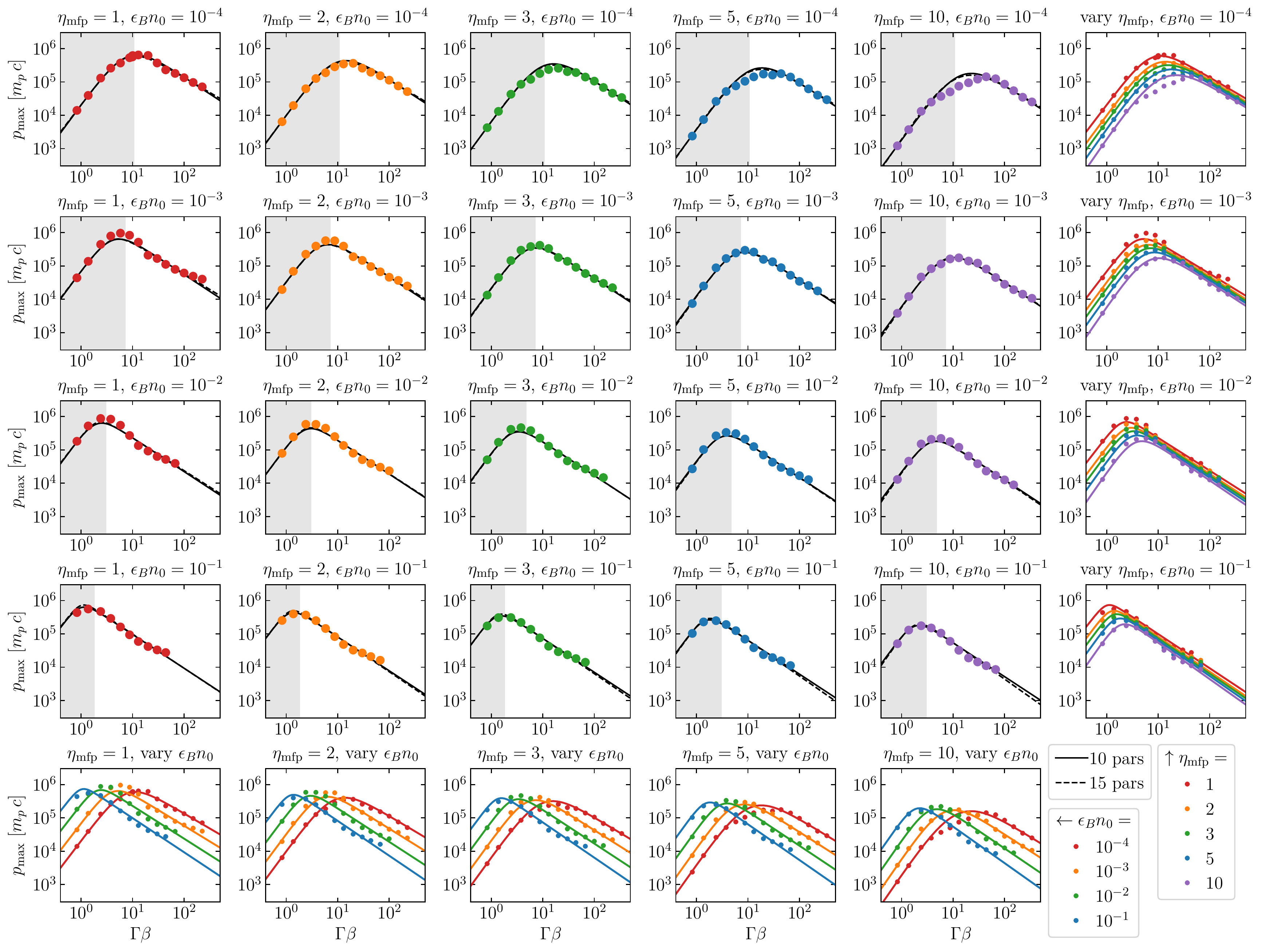}
  \caption{Dependence of Equation~\ref{eq:pl_def}'s $\pmax$ on $\gb$ for the various combinations of $\epsB n_{0}$ and $\etamfp$ simulated.  The curves correspond to the $\pmax$ values predicted by the empirical expression (Equations~\ref{eq:broken_pl_used}-\ref{eq:W_def}) in its full (15-parameter, dashed line) and reduced (10-parameter, solid line) forms; the values of the parameters are provided in Table~\ref{tab:param_sets} below.  Gray shading marks regions of the parameter space where electrons were able to escape from the shock in the upstream direction.  The rightmost column compares MC runs with the same $\epsB n_{0}$ but different $\etamfp$.  The bottom row compares MC runs with the same $\etamfp$ but different $\epsB n_{0}$.}
  \label{fig:mcmc_round1}
\end{figure*}
The values of $\pmax$ for each $\phiMC$ are shown in Figure~\ref{fig:mcmc_round1}.  While all four of Equation~\ref{eq:pl_def}'s $\pipl$ parameters are estimated for each $\phiMC$, hereafter we focus only on $\pmax$.  Each dot in Figure~\ref{fig:mcmc_round1} represents the highest-likelihood value of $\pmax$ from a single $\phiMC$ (cf. Figure~\ref{fig:mc_results}).  Within each subpanel, corresponding to a specific $(\epsB n_{0}, \etamfp)$ doublet, $\pmax$ versus $\gb$ appears to follow a broken power law.  The shaded area in each subpanel marks the region of $\gb$ values for which electrons were able to scatter upstream to the FEB discussed in Section~\ref{sec:code}.  It appears from Figure~\ref{fig:mcmc_round1} that electron escape is, at the very least, strongly correlated with the break in the $\pmax$ curves.  That is, below the break the maximum electron energy is limited by escape at the FEB, while above the break the limiting process is radiative losses.

\section{An Empirical Expression for $\pmax$}
\label{sec:fit}

As seen in Figure~\ref{fig:mcmc_round1}, $\pmax$ appears to follow a broken power law in $\gb$:
\begin{equation}
  \pmax(y = \gb) = A \left( \frac{y}{B} \right)^{C} \left[ \frac{1}{2} \left\lbrace 1 + \left( \frac{y}{B} \right)^{E} \right\rbrace \right]^{-\frac{C+D}{E}}
  \label{eq:broken_pl_base}
\end{equation}
In this form, the parameters $A$, $B$, $C$, $D$, and $E$ do not have well-defined physical associations.  For example, while setting $y = B$ results in $\pmax = A$, it is not the case that $B$ corresponds to the location of the peak and $A$ to the height at the peak. Instead, both the location and the height of the peak also depend on $C$, $D$, and $E$.  It is therefore useful to re-arrange Equation~\ref{eq:broken_pl_base} so that the five parameters describe different, mostly independent features of the broken power law.  This yields
\begin{equation}
  \pmax(y = \gb) = p_\mathrm{pk} \left( \frac{y}{\Gamma_\mathrm{pk}\beta_\mathrm{pk}} \right)^{L} \left[ \frac{1 + \frac{L}{H}}{1 + \left( \frac{L}{H} \right) \left( \frac{y}{\Gamma_\mathrm{pk}\beta_\mathrm{pk}} \right)^{W} } \right]^{ \frac{L+H}{W} }
  \label{eq:broken_pl_used}
\end{equation}
whose five features are $p_\mathrm{pk}$, the maximum value of the broken power law; $\Gamma_\mathrm{pk}\beta_\mathrm{pk}$, the value of $\gb$ at which the function peaks; $L$ and $H$, the asymptotic power-law indices far below and above (respectively) the peak; and finally $W$, which controls the width of the peak.

We seek an empirical formula for $\pmax$ over the three-dimensional $\phiMC = (\gb, \epsB n_{0}, \etamfp)$ parameter space.  Since Equation~\ref{eq:broken_pl_used} already explicitly depends on $\gb$, this can be accomplished by assuming that all five features of Equation~\ref{eq:broken_pl_used} depend on the remaining two parameters, $\epsB n_{0}$ and $\etamfp$, such that
\begin{equation}
  p_\mathrm{pk} = p_{C} \, \left(\etamfp\right)^{p_{\eta}} \, \left(\epsB n_{0}\right)^{p_{\epsilon}} m_{p} c
  \label{eq:ppk_def}
\end{equation}
\begin{equation}
  \Gamma_\mathrm{pk}\beta_\mathrm{pk} = g_{C} \, \left(\etamfp\right)^{g_{\eta}} \, \left(\epsB n_{0}\right)^{g_{\epsilon}}
  \label{eq:gbpk_def}
\end{equation}
\begin{equation}
  L = L_{C} \, \left(\etamfp\right)^{L_{\eta}} \, \left(\epsB n_{0}\right)^{L_{\epsilon}}
  \label{eq:L_def}
\end{equation}
\begin{equation}
  H = H_{C} \, \left(\etamfp\right)^{H_{\eta}} \, \left(\epsB n_{0}\right)^{H_{\epsilon}}
  \label{eq:H_def}
\end{equation}
\begin{equation}
  W = W_{C} \, \left(\etamfp\right)^{W_{\eta}} \, \left(\epsB n_{0}\right)^{W_{\epsilon}}
  \label{eq:W_def}
\end{equation}
The subscript $C$ here denotes the coefficient parameter for each feature, and the subscripts $\eta$ and $\epsilon$ identify the power parameters that quantify the dependence of each feature on $\etamfp$ and $\epsB n_{0}$ respectively.  Note that $p_\mathrm{pk}$ is the only feature with units: specifically, it is scaled to $m_{p} c$.

The PLDs for the 15 parameters in Equations~\ref{eq:ppk_def}-\ref{eq:W_def} were estimated via MCMC as described in Appendix~\ref{sec:MCMC_details}.  The results are presented in Table~\ref{tab:param_sets}.  The estimates suggest that $p_\mathrm{pk}$ depends only weakly on $\epsB n_{0}$, since $p_{\epsilon}$ is close to 0.  Similarly, $L$ and $H$ have a very weak dependence on both $\epsB n_{0}$ and $\etamfp$.  Therefore the parameter estimation procedure was repeated with a reduced 10-parameter expression, making the following modifications:
\begin{equation}
  p_\mathrm{pk} = p_{C} \left(\etamfp\right)^{p_{\eta}} m_{p} c
  \label{eq:ppk_reduced}
\end{equation}
\begin{equation}
  L = L_{C}
  \label{eq:L_reduced}
\end{equation}
\begin{equation}
  H = H_{C}
  \label{eq:H_reduced}
\end{equation}
Parameter estimates for the reduced 10-parameter expression also appear in Table~\ref{tab:param_sets}.

\begin{table}[h]
\centering
\begin{tabular}{lll}
\hline
Parameter & 15-parameter & 10-parameter \\
\hline
$p_{C}$ & $\phantom{-}{7.88\xx{5}}\,^{+3400}_{-3100}$ & $\phantom{-}{6.35\xx{5}}\,^{+1100}_{-1100}$ \\
$p_{\eta}$ & $-0.575\,^{+0.0015}_{-0.0018}$ & $-0.548\,^{+0.0014}_{-0.0015}$ \\
$p_{\epsilon}$ & $\phantom{-}0.0297\,^{+0.00048}_{-0.0005}$ & $\phantom{-}0$ \\
\hline
$g_{C}$ & $\phantom{-}0.527\,^{+0.002}_{-0.0018}$ & $\phantom{-}0.532\,^{+0.0012}_{-0.0012}$ \\
$g_{\eta}$ & $\phantom{-}0.299\,^{+0.0016}_{-0.0016}$ & $\phantom{-}0.278\,^{+0.00092}_{-0.001}$ \\
$g_{\epsilon}$ & $-0.333\,^{+0.00054}_{-0.00048}$ & $-0.338\,^{+0.00032}_{-0.00032}$ \\
\hline
$L_{C}$ & $\phantom{-}1.98\,^{+0.024}_{-0.023}$ & $\phantom{-}2.12\,^{+0.0074}_{-0.007}$ \\
$L_{\eta}$ & $\phantom{-}0.0464\,^{+0.0033}_{-0.0044}$ & $\phantom{-}0$ \\
$L_{\epsilon}$ & $-0.00346\,^{+0.0014}_{-0.0013}$ & $\phantom{-}0$ \\
\hline
$H_{C}$ & $\phantom{-}1.07\,^{+0.0029}_{-0.0029}$ & $\phantom{-}1.01\,^{+0.0012}_{-0.0012}$ \\
$H_{\eta}$ & $\phantom{-}0.0282\,^{+0.0014}_{-0.0013}$ & $\phantom{-}0$ \\
$H_{\epsilon}$ & $\phantom{-}0.0183\,^{+0.00049}_{-0.00048}$ & $\phantom{-}0$ \\
\hline
$W_{C}$ & $\phantom{-}7.7\,^{+0.13}_{-0.11}$ & $\phantom{-}5.26\,^{+0.039}_{-0.042}$ \\
$W_{\eta}$ & $-0.179\,^{+0.006}_{-0.0054}$ & $-0.043\,^{+0.0026}_{-0.0026}$ \\
$W_{\epsilon}$ & $\phantom{-}0.136\,^{+0.0019}_{-0.0018}$ & $\phantom{-}0.11\,^{+0.00073}_{-0.00068}$ \\
\hline
\end{tabular}
\caption{\label{tab:param_sets} The mode of each parameter's PLD; and the boundaries of the 68\% credible region of the PLD for each parameter, marginalized over all others. For the five fixed parameters of the 10-parameter set, no credible region is given.}
\end{table}

The parameter estimates in Table~\ref{tab:param_sets} suggest that when electron energy is limited by the location of the FEB, the value of $\pmax$ scales as roughly $(\gb)^{2}$; once radiation losses are the limiting factor, $\pmax$ scales as approximately $(\gb)^{-1}$.  Neither the value of $L$ nor that of $H$ depends strongly on the acceleration efficiency ($\etamfp$) or the peak magnetic field strength ($\epsB n_{0}$), motivating the reduction from a 15-parameter set to just 10.  The location of the break in the power law, $\Gamma_\mathrm{pk}\beta_\mathrm{pk}$, decreases as the magnetic field strength goes up, and increases as acceleration efficiency drops.  The height of the break, $p_\mathrm{pk}$, decreases as acceleration efficiency drops, and shows almost no dependence on the magnetic field strength.  As for the width parameter $W$, it increases with larger values of $\epsB n_{0}$ and with smaller values of $\etamfp$; since larger $W$ means a sharper peak, however, the trend is for broader breaks in the power law when either the magnetic field is weak or acceleration is inefficient.

With the values presented in Table~\ref{tab:param_sets}, it is possible to compute the value of $\pmax$ for any desired $\phiMC$.  First $\etamfp$ and $\epsB n_{0}$ are used in Equations~\ref{eq:ppk_def}--\ref{eq:W_def} to compute the five features of Equation~\ref{eq:broken_pl_used}.  The computed features are, in turn, used to obtain $\pmax$ by evaluating Equation~\ref{eq:broken_pl_used} at the desired value of $\gb$.

\section{Discussion}
\label{sec:disc}

We now consider two applications of Equation~\ref{eq:broken_pl_used} to GRB afterglows.  The first is the upper limit on synchrotron photon energy produced by the external shock of a GRB, and the second is a brief discussion of the physical process that limits electron energy in our simulations.

\subsection{Maximum energy of synchrotron photons}
\label{sub:syn_phot_max}

GRB afterglows are routinely associated with emission above 100 MeV, and analytical treatments of particle acceleration suggest that electron synchrotron photons can exceed $\sim$GeV energies while the forward shock of the GRB is still moving ultrarelativistically \citep{PiranNakar2010,Kumar_etal_2012}.  The MC simulations presented in Section~\ref{sec:results} serve as a numerical check on previous analytical calculations, as they take place under largely the same assumed conditions (c.f. Section~\ref{sec:code}).  We now calculate $h\nu_\mathrm{syn,max} = 3 \gamma_\mathrm{max}^{2} h q B / (4 \pi m_{e} c)$ for electron distributions whose maximum momentum $\pmax$ is given by Equation~\ref{eq:broken_pl_used}.  This is the characteristic synchrotron frequency of such electrons; the resultant photon spectrum would peak at a frequency approximately $0.3 \nu_\mathrm{syn,max}$ \citep{RybickiLightman1979}.

Let us now restrict our focus only to the portion of Equation~\ref{eq:broken_pl_used} above $p_\mathrm{pk}$, where radiative losses rather than escape act to limit the maximum electron energy.  At shock Lorentz factors far above $\Gamma_\mathrm{pk}\beta_\mathrm{pk}$, Equation~\ref{eq:broken_pl_used} simplifies to
\begin{equation}
  p_\mathrm{max} \approx p_\mathrm{pk} \left[ \frac{1 + \frac{L}{H}}{\frac{L}{H}} \right]^\frac{L+H}{W} \left( \frac{\Gamma}{\Gamma_\mathrm{pk}\beta_\mathrm{pk}} \right)^{-H} ,
  \label{eq:broken_pl_simpl}
\end{equation}
where we have reduced the independent variable to $\Gamma$ rather than $\gb$.  Using the 10-parameter values from Table~\ref{tab:param_sets}, the above equation can be expressed
\begin{equation}
  \gamma_\mathrm{max} = 9.4\xx{8} \, \etamfp^{-0.267} \epsB^{0.281} n_{0}^{0.281} \Gamma^{-1.01} ,
  \label{eq:elec_gam_max}
\end{equation}
having recast the equation in terms of Lorentz factor rather than momentum.  (The feature $W$ cannot be evaluated without knowing the value of $\epsB n_{0}$; the dependence on $\etamfp$ is weak enough to be ignorable.  For the range of conditions simulated in this work $W$ varies between 1.73 and 4.08.  We have taken 2.90 as a fiducial value to eliminate the dependence of $\gamma_\mathrm{max}$ on $W$.  The eventual result for $h \nu_\mathrm{syn,max}$ differs by a factor close to unity as a consequence of this simplification.)

We now assume that the large-scale hydrodynamics are governed by the \cite{BlandfordMcKee1976} solution for a relativistic impulsive explosion.  Further assuming an adiabatic shock, with neither energy injection nor significant radiative losses, one may derive the following expression for the shock Lorentz factor $\Gamma$:
\begin{equation}
  \Gamma^{2} = \left[ \frac{(17 - 4k) E_\mathrm{iso}}{8 \pi N m_{p} c^{5-k} \left[ 2(4-k)\right]^{3-k}} \right]^{\frac{1}{4-k}} t_\mathrm{obs}^{-\frac{3-k}{4-k}} .
  \label{eq:BM_gamma}
\end{equation}
In this equation, $k$ controls the density dependence of the circumburst medium (0 for constant density, and 2 for a wind-like medium); $E_\mathrm{iso}$ is the isotropic-equivalent explosion energy of the GRB; $N$ is the density parameter (such that the number density encountered by the shock is $n_{0} = N R^{-k}$ at some distance $R$); and $t_\mathrm{obs}$ is the time since the start of the GRB in the observer frame.  (Cosmological redshift may be significant, but we omit the factors of $(1+z)$ for the sake of simplicity.)  A similar application of the Blandford--McKee solution results in an expression for the ambient density $n_{0}$:
\begin{equation}
  n_{0} = N c^{-k} t_\mathrm{obs}^{-k} \Gamma^{-2k} \left[ 2(4-k) \right]^{-k}.
  \label{eq:BM_n0}
\end{equation}

The characteristic frequency of synchrotron emission depends on the electron Lorentz factor, $\gamma_\mathrm{max}$ as shown in Equation~\ref{eq:elec_gam_max}, and on the magnetic field strength $B$ as defined by Equation~\ref{eq:epsB_pk}.  We can specify $k = 0$ or $k = 2$ in these two equations, replacing $N$ with $n_\mathrm{ism}$ or $A_{\star}$ as appropriate:
\begin{equation}
  \gamma_\mathrm{max}^{2} = 
  \begin{cases}
    4.72\xx{18} \, \etamfp^{-0.534} \, \epsB^{-0.682} \, n_\mathrm{ism}^{-0.43} \, E_\mathrm{iso}^{-0.253} \, t_\mathrm{obs}^{0.758} \, m_{p}^{0.253} \, c^{1.26}  &  k = 0  \\
    3.79\xx{18} \, \etamfp^{-0.534} \, \epsB^{-0.682} \, A_{\star}^{-0.859} \, E_\mathrm{iso}^{0.177} \, t_\mathrm{obs}^{1.19} \, m_{p}^{-0.177} \, c^{0.833}  &  k = 2
  \end{cases}
  \label{eq:elec_gam_instantiated}
\end{equation}
\begin{equation}
  B = 
  \begin{cases}
    3.10 \, \epsB^{1/2} \, n_\mathrm{ism}^{3/8} \, E_\mathrm{iso}^{1/8} \, t_\mathrm{obs}^{-3/8} \, m_{p}^{3/8} \, c^{3/8}  &  k = 0  \\
    3.24 \, \epsB^{1/2} \, A_{\star}^{3/4} \, E_\mathrm{iso}^{-1/4} \, t_\mathrm{obs}^{-3/4} \, m_{p}^{3/4} \, c^{3/4}  &  k = 2
  \end{cases}
  \label{eq:B_pk_instantiated}
\end{equation}
The preceding two equations are sufficient to compute the maximum synchrotron frequency in the rest frame of the emitting plasma.  Boosting to the observer frame requires multiplying by the Doppler factor $\mathcal{D}$,
\begin{equation}
  \mathcal{D} \approx 2 \Gamma =
  \begin{cases}
    0.873 \, E_\mathrm{iso}^{1/8} \, n_\mathrm{ism}^{-1/8} \, t_\mathrm{obs}^{-3/8} \, m_{p}^{-1/8} \, c^{-5/8}  &  k = 0 \\
    1.09 \, E_\mathrm{iso}^{1/4} \, A_{\star}^{-1/4} \, t_\mathrm{obs}^{-1/4} \, m_{p}^{-1/4} \, c^{-3/4}  &  k = 2
  \end{cases} \\
  \label{eq:doppler_boost}
\end{equation}
Finally,
\begin{align}
  h \nu_\mathrm{syn,max} &= \frac{3 q h}{4 \pi m_{e} c}
  \begin{cases}
    1.26\xx{19} \, \etamfp^{-0.534} \,
    \epsB^{-0.182} \,
    n_\mathrm{ism}^{-0.18} \,
    E_\mathrm{iso}^{-0.003} \,
    t_\mathrm{obs}^{0.008} \,
    m_{p}^{0.503} \,
    c^{1.01}  &  k = 0 \\
    1.34\xx{19} \, \etamfp^{-0.534} \,
    \epsB^{-0.182} \,
    A_{\star}^{-0.359}
    E_\mathrm{iso}^{0.177}
    t_\mathrm{obs}^{0.187}
    m_{p}^{0.323}
    c^{0.833}  &  k = 2
  \end{cases} \nonumber \\
  &=
  \begin{cases}
    660 \, \mathrm{MeV} \, \etamfp^{-0.534} \, \epsB^{-0.182}
    \left( n_\mathrm{ism,0} \right)^{-0.18}
    \left( E_\mathrm{iso,53} \right)^{-0.003}
    \left( t_\mathrm{obs,2} \right)^{0.008}  &  k = 0 \\
    920 \, \mathrm{MeV} \, \etamfp^{-0.534} \, \epsB^{-0.182}
    \left( A_{\star,34} \right)^{-0.359}
    \left( E_\mathrm{iso,53} \right)^{0.177}
    \left( t_\mathrm{obs,2} \right)^{0.187}  &  k = 2
  \end{cases}
  \label{eq:syn_phot_max}
\end{align}
where we have used the notation $Q = Q_{x}\xx{x}$, e.g. $A_{\star} = A_{\star,34}\xx{34}$~cm$^{-1}$, and all quantities use cgs units.  We remind the reader that the observed peak of the photon distribution would occur at energies roughly 30\% of those given in Equation~\ref{eq:syn_phot_max}.

In the $k = 0$ case, the exponents of energy and time suggest that they may be due to uncertainty in the parameter $H_{C}$, and that the value of $h\nu_\mathrm{syn,max}$ may not depend on either quantity: if $H_{C}$ were precisely 1 instead of 1.01, neither factor would appear in Equation~\ref{eq:syn_phot_max}.  That is, the results presented in Figure~\ref{fig:mcmc_round1} suggest that the maximum observed synchrotron energy is essentially independent of both explosion energy and time.  For the $k = 2$ case, the maximum synchrotron frequency grows with time; though both Doppler factor and peak magnetic field decay with time, they are more than compensated for by the growth in maximum electron Lorentz factor.

The above limits should hold as long as (1) the motion of the blast wave is given by the adiabatic Blandford--McKee solution, (2) the shock Lorentz factor is above $\Gamma_\mathrm{pk}\beta_\mathrm{pk}$ given by Equation~\ref{eq:gbpk_def}, and (3) there is amplified microturbulence that decays on either side of the shock as prescribed by Equation~\ref{eq:epsB_form}.  Condition (1) means that Equation~\ref{eq:syn_phot_max} may not apply very early on in the afterglow, if the shock is still in the acceleration or coasting phases of its expansion. Condition (2) is satisfied for $\gb \gsim 10$.  Using this limitation in Equation~\ref{eq:BM_gamma} leads to the following restriction on $t_\mathrm{obs}$:
\begin{equation}
  t_\mathrm{obs} \lesssim
  \begin{cases}
    3.2\xx{4} \, \mathrm{s} \left( E_\mathrm{iso,53} \right)^{1/3} \left( n_\mathrm{ism,0} \right)^{-1/3}  &  k = 0 \\
    2\xx{6} \, \mathrm{s} \, E_\mathrm{iso,53} \, \left( A_{\star,34} \right)^{-1}  &  k = 2
  \end{cases}
  \label{eq:tobs_limit}
\end{equation}

As shown above, both ISM and wind media allow for electron synchrotron photons that reach GeV energies; but the extremely high energy photons detected in events like GRB 090902B \citep[33.4 GeV;][]{Abdo_etal_2009ApJ706}, GRB 130427 \citep[95 GeV;][]{Ackermann_etal_2014}, and GRB 190114C \citep[$>300$ GeV;][]{Mirzoyan2019ATel12390} suggest that an additional channel produces the highest-energy photons detected in GRB afterglows.  \citep[Alternately, electron acceleration could proceed in a more exotic scenario than modeled in this work, e.g.][who consider acceleration in the presence of clumpy magnetic field condensations.]{Khangulyan_etal_2020}  Though thermal electrons were ignored in this work, the synchrotron self-Compton mechanism in this population is a straightforward way to produce emission that peaks in the GeV--TeV range well into the afterglow \citep{WEBN2017}.

\subsection{What process limits electron energy?}
\label{sub:process}

The idea of using the synchrotron cooling break to interpret afterglow observations is nearly as old as afterglow observations themselves \citep{SPN1998}; for a recent example of a work using this formalism, see \citet{Fraija_etal_2020}, or the numerous citations within.  There are multiple approaches to determining the location of the cooling break.  One method is to assume that the electron distribution immediately behind the shock extends to infinity \citep{GranotSari2002}; synchrotron losses then quickly cool the highest-energy electrons and limit the extent of the electron distribution downstream from the shock.  Another, more realistic, way is to equate the electron acceleration time to the synchrotron cooling time to determine the energy at which acceleration stops \citep{Lemoine2013,Khangulyan_etal_2020}.

None of the previous analytical methods would seem to apply to the acceleration scenario presented in this work.  For example, consider the simplest case: electrons scattering in a homogeneous magnetic field.  In this setting, the synchrotron loss time is given by
\begin{equation}
  t_\mathrm{syn} = \frac{ 5 \xx{8} }{ B^{2} \, \gamma_\mathrm{elec} }~\mathrm{s} \propto B^{-2} \, \gamma_\mathrm{elec}^{-1}
  \label{eq:syn_loss_time}
\end{equation}
for electrons with energy $\gamma_\mathrm{elec}$ moving in a homogeneous field of strength $B$.  The maximal electron energy may be calculated by comparing the above loss time to the typical residence time of an electron in the downstream region of the shock. In keeping with the Bohm approximation above, we assume that the residence time is proportional both to the gyroperiod of the electron and to the Bohm factor:
\begin{equation}
  t_\mathrm{res|d} \propto \frac{ \etamfp \, 2 \pi r_{g} }{ c } \propto \etamfp \, \gamma_\mathrm{elec} \, B^{-1} .
  \label{eq:tres_DwS}
\end{equation}
For electrons with the maximum achievable energy, $t_\mathrm{syn}$ and $t_\mathrm{res|d}$ should be roughly equal, yielding
\begin{equation}
  B^{-2} \gamma_\mathrm{max}^{-1} \propto \etamfp \gamma_\mathrm{max} B^{-1} .
  \label{eq:gammax_prop_1}
\end{equation}
The magnetic field is determined by Equation~\ref{eq:B_pk_strength}.  Substituting that equation into Equation~\ref{eq:gammax_prop_1} leads to the proportionality
\begin{equation}
  \gamma_\mathrm{max} \propto \etamfp^{-1/2} n_{0}^{-1/4} \Gamma_{0}^{-1/2} \epsB^{-1/4}
  \label{eq:gammax_prop_2}
\end{equation}
Equation~\ref{eq:gammax_prop_2} has a different dependence on $\Gamma_{0}$ ($-1/2$) than displayed in Figure~\ref{fig:mcmc_round1} and Table~\ref{tab:param_sets} ($\approx -1$).  This suggests that the above scenario does not reflect the energy-limiting process occurring in the MC simulations.  This is unsurprising, as Equations~\ref{eq:syn_loss_time} and \ref{eq:tres_DwS} assumed a homogenous magnetic field of strength $B$, while the MC simulations took place in a decaying microturbulent magnetic field.

Another straightforward limit on the maximum attainable energy of a shock-accelerated particle is to equate the acceleration time to the dynamical time of the shock.  The former quantity is conservatively given by the downstream residence time in Equation~\ref{eq:tres_DwS}.  The latter is proportional to $t_\mathrm{shock}/\Gamma$ for a shock of age $t_\mathrm{shock}$.  Setting the two to be roughly equal leads to the proportionality
\begin{equation}
  \gamma_\mathrm{max} \propto \etamfp^{-1} n_{0}^{1/6} \GamZ^{-2/3} \epsB^{1/2}, 
  \label{eq:gammax_tacc_tdyn}
\end{equation}
where we have substituted Equation~\ref{eq:B_pk_strength} into Equation~\ref{eq:tres_DwS}.  This approach also fails to reproduce the dependence on $\Gamma$ seen in Figure~\ref{fig:mcmc_round1} and Table~\ref{tab:param_sets}, suggesting either that the assumptions made are not correct or that the maximum electron energy is not limited by time.

The broad structure visible in Figure~\ref{fig:mcmc_round1} is that of a broken power law, which suggests that two processes control $\pmax$ in our model of electron acceleration.  There are hints, however, of more structure---particularly at the top right corner (high $\etamfp$ and low $\epsB n_{0}$), where the empirical equation seems to slightly overestimate $\pmax$, and at the bottom left corner (low $\etamfp$ and high $\epsB n_{0}$), where there may be an additional power law segment at higher $\gb$ values.  These possible additional features could be physical in origin, something we intend to investigate in future work.

\section{Conclusions}
\label{sec:concl}

We have presented here the results of numerous Monte Carlo (MC) simulations of diffusive shock acceleration of electrons.  Acceleration took place in a highly turbulent magnetic field that decayed upstream and downstream from the shock, as observed in particle-in-cell simulations of relativistic shocks.  The accelerated electron spectra are well-described by power laws with an exponential turnover (Equation~\ref{eq:pl_def}, Figure~\ref{fig:mc_results}), and the likelihood distributions of their parameters $\pipl = (\pmax, \sigma, \chi, \log_{10}K)$ were estimated using a Markov chain Monte Carlo (MCMC) approach.  We focused here only on the parameter $\pmax$, i.e. the location of the exponential turnover and the energy at which acceleration gains balance loss processes.  When the estimates of $\pmax$ are plotted across a 3-dimensional parameter space of astrophysical interest, they appear to form broken power laws as a function of the shock speed $\gb$ (Figure~\ref{fig:mcmc_round1}).  The peak in each curve lies close to the shock speed at which electrons cease to escape upstream from the shock, which suggests that upstream escape and radiation losses are the two limiting processes that control the maximum electron energy.

Based on the broken power law observed for $\pmax$ as a function $\gb$, we proposed an empirical expression in terms of physically-motivated features (Equation~\ref{eq:broken_pl_used}).  Another round of MCMC was used to estimate its parameters (Table~\ref{tab:param_sets}).  The resulting expression allows for prediction of the peak electron energy---and correspondingly the synchrotron cooling energy---in scenarios where relativistic shocks accelerate electrons, without the need for time-consuming numerical simulations.  Extensions to this work will involve increasing the size and dimensionality of the MC-simulated parameter space $\phiMC = (\etamfp, \epsB n_{0}, \gb)$ from what was considered here, as well as quantitative treatments of the $\sigma$ and $\chi$ parameters in Equation~\ref{eq:pl_def}.

\appendix

\section{Parameter estimation via MCMC}
\label{sec:MCMC_details}

\subsection{Estimating electron spectrum parameters $\pipl$}

Throughout this work, posterior likelihood distributions (PLDs) for estimated parameters are obtained via Markov chain Monte Carlo (MCMC) using the \texttt{phymcmc} \cite{phymcmc} python package, which makes use of \texttt{emcee} \cite{Foreman-Mackey_etal_2013}.

In Section~\ref{sec:results}, the PLD of the dependent parameters $\pipl = (\pmax, \sigma, \chi, \log_{10}K)$ of Equation~\ref{eq:pl_def} given the spectra generated by the Monte Carlo (MC) model for each $\phiMC = (\etamfp, \epsB n_{0}, \gb)$ was computed as
\begin{equation}
  \text{PLD}(\pipl|\mathrm{data}) =
\frac{\mathcal{P}(\mathrm{data}|\pipl) \cdot \text{Prior}(\pipl)}{\mathcal{P}(\text{data})} 
\propto \exp\left[-\frac{\mathrm{SSR}(\mathrm{data},\pipl)}{2\sigma^2(\mathrm{data})}\right] \cdot \text{Prior}(\pipl) \ ,
\end{equation}
where
\begin{equation}
  \frac{\mathrm{SSR}(\mathrm{data},\pipl)}{\sigma^2(\mathrm{data})} = \sum_{p_i} \frac{\left\{\mathrm{data}(p_i,\phiMC) - \log_{10}[\mathrm{eqn}(p_i,\pipl)]\right\}^2}{\text{variance}\{\mathrm{data}(p_i,\phiMC)\}} \ ,
\end{equation}
with $\mathrm{data} = \log_{10}[\dNdp](p_i,\phiMC)$, the $\log_{10}$ particle density at momentum $p_i$ for each of the 20 iterations of the MC code, $\text{variance}\{\mathrm{data}(p_i,\phiMC)\}$ is the variance in the data computed at each $p_i$, and $\mathrm{eqn}(p_i,\pipl)$ is the value of Equation~\ref{eq:pl_def} at $p_i$ given $\pipl$. A log-uniform prior was assumed for all $\pipl$ parameters such that
\begin{equation}
  \mathrm{Prior}(\pipl) \propto \frac{1}{\pmax \cdot\ \sigma \cdot\ \chi \cdot\ \log_{10}K}\ .
\end{equation}
After a sufficient burn-in, the PLDs were obtained from 300 chains of 6,000 steps each, which yielded $\sim$45,000 independent $\pipl$ sets. The resultant PLD for each $\pipl$ parameter, marginalized over all other $\pipl$ parameters, were all log-normal distributions, for each of $\pmax$, $\sigma$, $\chi$ and $\log_{10}(K)$.

\subsection{Estimating the parameters of the empirical equation, $\pemp$}

In Section~\ref{sec:fit}, the log-normal distributions for $\pmax$ (obtained in Section~\ref{sec:results} for a wide range of $\phiMC$ values) were used to estimate the 15 parameters, $\pemp$, of the empirical expression (Equations~\ref{eq:broken_pl_used}--\ref{eq:W_def}) that computes $\pmax$ directly from $\phiMC$. The PLD of $\pemp$ is computed as
\begin{equation}
  \text{PLD}(\pemp|\mathrm{data}) = \frac{\mathcal{P}(\mathrm{data}|\pemp) \cdot \text{Prior}(\pemp)}{\mathcal{P}(\text{data})} \propto \exp\left[-\frac{\mathrm{SSR}(\mathrm{data},\pemp)}{2\sigma^2(\mathrm{data})}\right] \cdot \text{Prior}(\pemp) \ ,
\end{equation}
where
\begin{equation}
  \frac{\mathrm{SSR}(\mathrm{data},\pemp)}{\sigma^2(\mathrm{data})} = \sum_{\phiMC} \frac{\left\{\mu(\phiMC) - \log[\mathrm{eqn}(\phiMC,\pemp)]\right\}^2}{\sigma^2(\phiMC)}\ ,
\end{equation}
with $\mu(\phiMC)$ and $\sigma^2(\phiMC)$ the mean and variance of the normally distributed posterior of $\ln(\pmax)$ at $\phiMC$, and $\mathrm{eqn}(\phiMC,\pemp)$ is the value of $\pmax$ computed from Equations~\ref{eq:broken_pl_used}--\ref{eq:W_def} given $\pemp$ and $\phiMC$. A linear uniform prior was assumed for all power parameters (those with subscripts $\epsilon$ and $\eta$), and a log-uniform prior was assumed for all coefficient parameters (subscript $C$), such that
\begin{equation}
  \mathrm{Prior}(\pemp) \propto \frac{1}{p_C \cdot\ g_C \cdot\ L_C \cdot\ H_C \cdot\ W_C}\ .
\end{equation}
After a sufficient burn-in, the PLDs were obtained from 700 chains of 6,000 steps each, which yielded $\sim$56,000 independent $\pemp$ sets.

\acknowledgments 
This work was funded in parts by the Interdisciplinary Theoretical and Mathematical Sciences (iTHEMS, ithems.riken.jp) program at RIKEN (DCW, CAAB, SN); and by Discovery Grant 355837-2013 from the Natural Sciences and Engineering Research Council of Canada (www.nserc-crsng.gc.ca, CAAB). MB would like to acknowledge support by NASA grants 80NSSC17K0757 and NNH19ZDA001N-FERMI, and NSF grants 10001562 and 10001521. This work is supported by JSPS Grants-in-Aid for Scientific Research “KAKENHI”(A: Grant Number JP19H00693). SN also acknowledges the support from Pioneering Program of RIKEN for Evolution of Matter in the Universe (r-EMU).

\bibliographystyle{aasjournal} 
\bibliography{dcw}{}

\end{document}